\documentclass[aps, prd, preprint, onecolumn, amsmath, amssymb, superscriptaddress,titlepage]{revtex4-1}
\usepackage[T1]{fontenc}
\usepackage{fancyhdr}
\usepackage{mathbbol}
\usepackage{amssymb}
\usepackage{bbm}
\usepackage{ragged2e}
\usepackage[us,12hr]{datetime}
\usepackage{graphicx}
\usepackage{amssymb}
\usepackage{mathrsfs}
\usepackage{marvosym}
\newcommand{\comment}[1]{}
\usepackage{booktabs}
\usepackage{graphicx}
\usepackage[english]{babel}
\usepackage[utf8]{inputenc}
\usepackage{graphicx}
\usepackage{color}
\usepackage{bbold} 
\usepackage{xcolor}
\usepackage{mathrsfs}
\usepackage[braket, qm]{qcircuit}
\usepackage[colorlinks=true
,urlcolor=mn
,anchorcolor==,citecolor=magenta
,filecolor=navyblue
,linkcolor=rindou2
,menucolor=navyblue
,pagecolor=navyblue
,linktocpage=true
,pdfproducer=medialab]{hyperref}
\usepackage{slashed}

\usepackage{aas_macros,empheq,fancybox,graphicx,multirow}
\usepackage{subfig}
\usepackage[normalem]{ulem}
\usepackage{hyperref}
\usepackage{graphicx}
\usepackage{amsmath}
\usepackage{amsthm}
\usepackage{amssymb}
\usepackage{xcolor}
\usepackage{mathrsfs}
\theoremstyle{plain}
\usepackage{amsfonts}

\definecolor{pc1}{rgb}{0.69, 0.25, 0.21}

\newcommand\norm[1]{\left\lVert#1\right\rVert}
 
\definecolor{rindou1}{rgb}{0.4431,0.2862,0.7960}
\definecolor{rindou2}{rgb}{0.0078,0.1215,0.4392}
\definecolor{lapis}{rgb}{0.0.0470,0.2941,0.5568}
\definecolor{mn}{rgb}{0.15, 0.35, 0.95}

\begin{document}
\title{Hamiltonian simulation of minimal holographic sparsified SYK model}

\author{Raghav G. Jha}
\email{raghav.govind.jha@gmail.com}
\affiliation{Thomas Jefferson National Accelerator Facility, Newport News, Virginia 23606, USA}
\vspace{30mm}

\begin{abstract}
The circuit complexity for Hamiltonian simulation of the sparsified SYK model with $N$ Majorana fermions and $q=4$ (quartic interactions), 
which retains holographic features (referred to as `minimal holographic sparsified SYK')  with $k\ll N^{3}/24$ (where $k$ is the total number of interaction terms times 1/$N$) 
using the second-order Trotter method and Jordan-Wigner encoding is found to be $\widetilde{\mathcal{O}}(k^{\alpha}N^{3/2} \log N (\mathcal{J}t)^{3/2}\varepsilon^{-1/2})$ 
where $t$ is the simulation time, $\varepsilon$ is the desired error in the implementation of the unitary $U = \exp(-iHt)$ measured by the operator norm, $\mathcal{J}$ is the disorder strength, and constant $\alpha < 1$. 
This complexity implies that with less than 100 logical qubits and about $10^{6}$ gates, it might be possible to achieve an advantage in this model and simulate real-time dynamics. 
\end{abstract}

\preprint{JLAB-THY-24-4027}
\maketitle
\section{Introduction}
The quest to find simple quantum many-body systems with holographic behavior is an important research direction that \emph{might} may also be helpful to perform lab experiments and understand the effects of quantum gravity. It would have been better if this quantum gravity were classically Newtonian gravity, but it still has merit as a toy model. The simplest quantum model (yet known)
with such a feature is the Sachdev-Ye-Kitaev (SYK) model. As we move towards the era of quantum computing, we would like to explore the possibility of studying these holographic models using quantum computers. 
We are far from this at the moment, as there is no problem in quantum many-body physics 
that \emph{can be} studied on a quantum computer but,  \emph{cannot} be matched 
using a classical computer. Most likely, this is expected to change in the next decade. 
Due to its importance, one would like to study the SYK model and compute observables
with minimum resources and go beyond the regime yet accessed using classical computers. One such observable is the out-of-time-order (OTOC) correlation function used in the study of quantum chaos. 
For the computation of this observable, we need to implement the time evolution, whose complexity is partially
determined by the structure of the Hamiltonian. The SYK model with all-to-all quartic
interaction results in a dense and nonlocal Hamiltonian. Although this is not \emph{very hard} to study with quantum
computers, a natural question is to what extent the model can be sparsified (that is, how many terms in the dense Hamiltonian can be ignored and set to zero) to maintain holographic behavior
such that it becomes slightly \emph{easier}. This was first studied in Refs.~\cite{Xu:2020shn, Garcia-Garcia:2020cdo} and
recent analysis \cite{Orman:2024mpw} suggests that
one can go pretty far (but not below some $k_{\text{critical}}$) in sparsifying the model and still have signs of holographic behavior
in the low-temperature ($\beta \mathcal{J} \gg 1$) limit. In this paper, we estimate the resources required to simulate this model in the noisy and fault-tolerant era and find that the dependence of resources on sparsity is sublinear in $k$.  
We find the leading order circuit complexity using the second-order Trotter formula is $\widetilde{\mathcal{O}}(k^{\alpha<1}N^{3/2} \log N (\mathcal{J}t)^{3/2}\varepsilon^{-1/2})$
with Jordan-Wigner (JW) mapping of the fermions to qubits. This can be improved to $\widetilde{\mathcal{O}}(k^{\alpha<1}N \log^{2} N (\mathcal{J}t)^{3/2}\varepsilon^{-1/2})$ 
using Bravyi-Kitaev (BK) encoding; however, we focus only on JW mapping.

\section{SYK Hamiltonian and the sparsified version} 

The Hamiltonian for the dense SYK model \cite{Maldacena:2016hyu} is given by: 
\begin{align}
H &= \frac{1}{4!} \sum_{a,b,c,d = 1}^{N} J_{abcd} \chi_{a} \chi_{b} \chi_{c} \chi_{d},
\label{eq:SYK_main} 
\end{align}
where $\chi$ are the Majorana fermions satisfying $\{\chi_a, \chi_b\} = 2\delta_{ab}$, $\chi_{i}^2 = \mathbb{1}$. 
$N$ Majorana fermions can be represented by $N/2$ fermions (complex), which have a local
two-dimensional Hilbert space and hence can be represented by one qubit. Therefore, for SYK
with $N$ Majorana fermions, we need $n = N/2$ qubits with the Hamiltonian, $H$, of size $2^{n} \times 2^{n}$.
The random disorder coupling $J_{abcd}$
is taken from a Gaussian distribution with mean
$\overline{J_{abcd}}=0$ and variance equal to
$\overline{J_{abcd}^{2}} = 6\mathcal{J}^2/N^3$ \cite{Maldacena:2016hyu}
where $\mathcal{J}$ has dimension of energy. 
For this dense $H$ with exactly four fermions (quartic) interacting in each term of the Hamiltonian, there is a total of $\binom{N}{4}$ terms
which grow like $\sim N^{4}/4!$ for large $N$. In previous work, we found the gate complexity for implementing time evolution using first-order Trotter (for fixed $t$ and error $\varepsilon$)
scaled like $\sim N^5$ unlike $\sim N^{10}$ discussed in Ref.~\cite{Garcia-Alvarez:2016wem}. For interesting holographic interpretation
and computation of out-of-time order correlators (OTOC), we need a large $N$ limit and sufficiently many Trotter steps to ensure that the Lyapunov exponent can be reliably computed. 
Therefore, we seek a simplified version of the SYK model such that the complexity can be further improved. 

Such a model is known as the `sparse SYK' model and was 
first discussed in \cite{Garcia-Garcia:2020cdo,Xu:2020shn}. The average number of terms in the Hamiltonian is $p \binom{N}{4}$.
The average degree of interaction hypergraph $k$ (defined as the number of hyperedges divided by the number of vertices, i.e., $N$) in the large $N$ limit is $p \frac{N^4}{24} \frac{1}{N} = \frac{pN^{3}}{24}$. If we take $k = \frac{N^3}{24}$, then $p=1$ and we have the dense (or the usual) SYK model. The minimum value of $k$ such that one can still have holographic features in the model is an open question, and the most recent estimate argues that $k \ge 8.7$ \cite{Orman:2024mpw}, while it was argued in \cite{Xu:2020shn} 
that a  smaller value i.e., $k \approx 4$ would also suffice. 
We only derive the Trotter
complexity of this model and do not apply it to compute any observable. 
For interesting simulations on IBM devices for the SYK model,
we refer the reader to Ref.~\cite{Asaduzzaman:2023wtd}.  
The Hamiltonian of the sparsified SYK model is: 
\begin{align}
H &= \frac{1}{4!} \sum_{a,b,c,d = 1}^{N} p_{abcd} J_{abcd} \chi_{a} \chi_{b} \chi_{c} \chi_{d},
\label{eq:SYK_sparse} 
\end{align}
where now we have an altered variance equal to 
$\overline{J_{abcd}^{2}} = \frac{6\mathcal{J}^2}{pN^3}$. Note that due to the random number that determines the removal of terms (i.e., sparsifying with some probability), the average number of terms is not fixed like the dense SYK but varies, and, therefore, the corresponding gate costs. For the dense model, $p_{abcd} = 1$ for all $\{a,b,c,d\}$. The terms in sparse SYK are removed with probability ($1-p$) and retained with probability $p$, i.e., $p_{abcd} = 0$ if a random uniform number between 0 and 1 is more  than $p$. This drastically alters the sparseness of the Hamiltonian and makes it \emph{easier} for quantum simulations. To quantum simulate this model, we use the standard Jordan-Wigner mapping to qubits where one needs $N/2$ qubits to describe 
a model with $N$ Majorana fermions. Using the 2$\times$2 Pauli matrices and identity matrix $\mathbb{1}$, we obtain:
\begin{align}
\chi_{2r-1} &= Z_{1} \cdots Z_{r-1} X_{r} \mathbb{1}_{r+1}\cdots  \mathbb{1}_{N/2},  \nonumber \\
\chi_{2r} &= Z_{1} \cdots Z_{r-1} Y_{r} \mathbb{1}_{r+1}\cdots  \mathbb{1}_{N/2},  \nonumber \\
\end{align}
where $k$ runs from $1, \cdots, N/2$ and the Kronecker product between 2$\times$2 matrices is not written explicitly. Once the Hamiltonian is written in terms of Pauli operators, we use the Trotter approach \cite{Seth1996} to simulate the model. Though popular, 
the Trotter method is not the only method to do the time evolution of a quantum system. Many better algorithms exist with the additional requirements of constructing specific 
oracles. For example, the Hamiltonian simulation algorithm in the fault-tolerant era based on qubitization argues that complexity to leading order for the 
dense SYK model is $\widetilde{\mathcal{O}}(N^{7/2}t)$ \cite{Babbush:2018mlj}. This 
is several orders of magnitude less than the first proposal using Trotter-based 
approach of Ref.~\cite{Garcia-Alvarez:2016wem} using $\sim N^{10}$ gates per step. 
This was recently improved to $\sim N^{5}$ in Ref.~\cite{Asaduzzaman:2023wtd}
using the commuting clusters of terms in the Hamiltonian and evolving the cluster together
by its diagonalizing circuit. In this work, we will only consider the Trotter-based methods due to their 
inherent simplicity and non-requirement of ancilla qubits or specific oracles.
To do such a Hamiltonian simulation, one decomposes the sparse SYK Hamiltonian into strings of Pauli operators 
as $H = \sum_{j=1}^{m} H_{j}$ where each $H_{j}$ is a tensor product of Pauli operators. Then using the first-order product formula (Trotter1) 
we can bound the error as follows: 
\begin{equation}
\bigg\vert\bigg\vert e^{-iHt} -  \Big(\prod_{j=1}^{m} e^{-iH_{j}t/r}\Big)^{r} \bigg\vert\bigg\vert \le \frac{\mathcal{J}^{2}t^2}{2r}
\sum_{p=1}^{m} \bigg\vert\bigg\vert \Big[ \sum_{q=p+1}^{m} H_q, H_p\Big] \bigg\vert\bigg\vert,
\label{eq:trotter_error1}
\end{equation}
\begin{figure}
    \centering
    \includegraphics[width=0.6\linewidth]{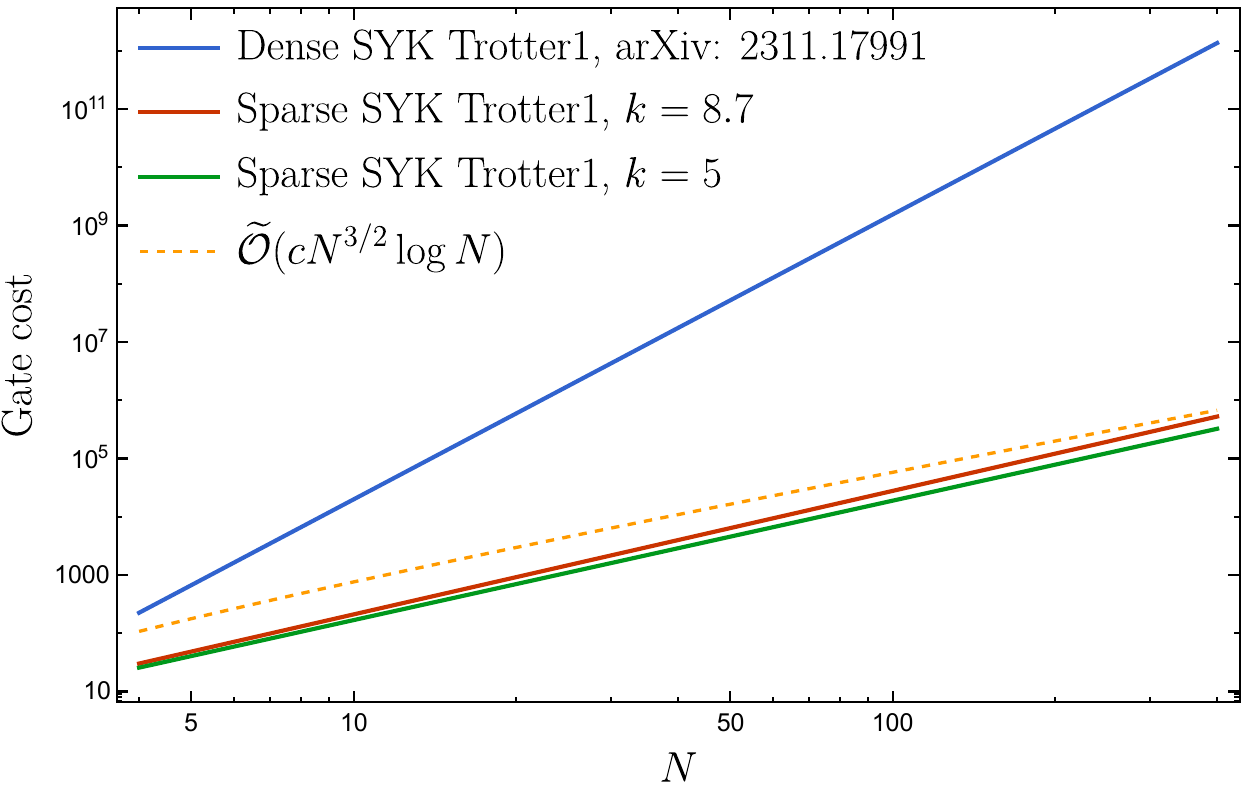}
    \caption{The circuit complexity to simulate the sparse SYK model with two choices of sparsification of the model per Trotter
    step. The gate costs are evaluated based on arranging the terms in the Hamiltonian in commuting clusters and 
    evolving each cluster by finding its diagonalizing circuit \cite{Asaduzzaman:2023wtd}. We denote the 
    sub-linear dependence on $k$ with $ \sim k^{\alpha}$ where $\alpha < 1$. We find that leading circuit complexity
    per Trotter step is $\widetilde{\mathcal{O}}(k^{p}N^{3/2} \log N$).}
    \label{fig:fig1}
\end{figure}
where we denote the unitary invariant spectral norm by $\norm{\cdots}$ and $m, r$ denote the 
number of terms in the decomposed Hamiltonian (less than the total number of terms since we cluster several terms together)
and the number of Trotter steps. Depending on the accuracy, second-order Trotter 
is also used in which case the error bounds are given by \cite{2019arXiv191208854C}:
\begin{align}
\bigg\vert\bigg\vert e^{-iHt} -  \Big(\prod_{j=m}^{1} e^{-iH_{j}t/2r} \prod_{j=1}^{m} e^{-iH_{j}t/2r} \Big)^{r} & \bigg\vert\bigg\vert \le \frac{\mathcal{J}^{3}t^3}{12r^{2}} \sum_{p=1}^{m} \bigg\vert\bigg\vert \Big[\sum_{r=p+1}^{m}H_{r}, \Big[\sum_{q=p+1}^{m} H_q, H_p\Big]\Big] \bigg\vert\bigg\vert \nonumber \\ 
& + \frac{\mathcal{J}^{3}t^3}{24r^{2}} \sum_{p=1}^{m} \bigg\vert\bigg\vert \Big[H_{p},  \Big[H_{p}, \sum_{r=p+1}^{m} H_r\Big]\Big] \bigg\vert\bigg\vert. 
\label{eq:trotter_error2}
\end{align}
For example, if we assume that $m$ is reasonably small compared to the total number of terms in the Hamiltonian as discussed in \cite{Asaduzzaman:2023wtd} and assuming that the operator norm is bounded by unity (this can be done by redefining the 
$H$ by including a factor of $1/N$), then the second-order product formula reads: 
\begin{equation}
   \bigg\vert\bigg\vert e^{-iHt} -  \Big(\prod_{j=m}^{1} e^{-iH_{j}t/2r} \prod_{j=1}^{m} e^{-iH_{j}t/2r} \Big)^{r} \bigg\vert\bigg\vert \le \mathcal{O}(\mathcal{J}^{3}t^{3}/r^{2}). 
\end{equation}
If we want to bound this error by $\varepsilon$, then we would need
to perform about $(\mathcal{J}t)^{3/2}/\sqrt{\varepsilon}$ Trotter steps.
We show the cost per Trotter step in terms of two-qubit gates and the cost in fault-tolerant era gates i.e., 
Clifford and $T$-gates in Table~\ref{tab:table1}
for $N \le 250$ for various $N$. 
The complexity using Jordan-Wigner encoding for
first-order Trotter is 
$\widetilde{\mathcal{O}}(k^{\alpha<1}N^{3/2} \log N (\mathcal{J}t)^{2} \varepsilon^{-1})$ 
while for second-order Trotter is $\widetilde{\mathcal{O}}(k^{\alpha<1}N^{3/2} \log N (\mathcal{J}t)^{3/2}\varepsilon^{-1/2})$.

\begin{figure}
    \centering
    \includegraphics[width=0.6\linewidth]{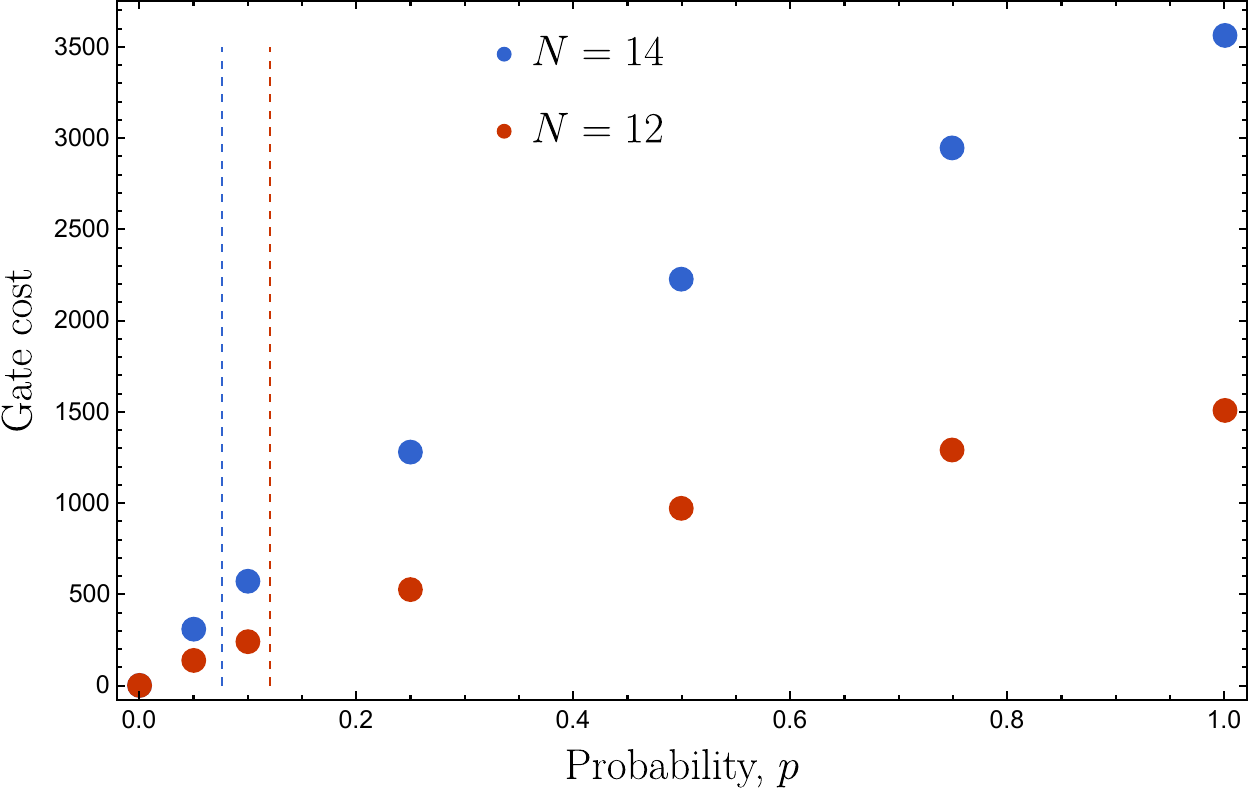}
    \caption{The dependence of circuit complexity on probability $p$ for $N=12,14$ SYK for each Trotter step.
    The standard SYK model with no pruning is at $p=1$. As we remove the terms randomly from the dense Hamiltonian, 
    the complexity decreases. The dashed lines roughly denote the minimum $p$ that retains holographic features for $N=12,14$.}
    \label{fig:fig2}
\end{figure}

\begin{table}[h]
\renewcommand{\arraystretch}{1.15}
\setlength{\tabcolsep}{12pt}
\centering
\begin{tabular}{|p{1cm}|p{2.5cm}|p{2.5cm}|p{2.5cm}|p{2.5cm}}
\hline
$N$ & CNOT gates & H+CNOT+ $T$ gates & $T$-gates \\ 
\hline
6 &  29 $\pm$ 1 & 126 $\pm$ 3 & 50  $\pm$ 2 \\
\hline
8 & 74  $\pm$ 2 & 349 $\pm$ 12 & 150  $\pm$ 7 \\ 
\hline
10 & 165 $\pm$ 5 & 645 $\pm$ 12 & 253  $\pm$ 6 \\
\hline
12 & 294 $\pm$ 6 & 1110 $\pm$ 22 & 430 $\pm$ 11 \\
\hline
14 & 440 $\pm$ 10 & 1593 $\pm$ 30 & 609 $\pm$ 12 \\
\hline
16 & 596 $\pm$ 16 & 2053 $\pm$ 56  & 763 $\pm$ 23 \\
\hline
18 & 771 $\pm$ 27 & 2478 $\pm$ 91 & 880 $\pm$ 38\\
\hline
20 & 967 $\pm$ 18 & 3097 $\pm$ 77 & 1104 $\pm$ 38 \\ 
\hline
30 & 2528 $\pm$ 77 & 6955 $\pm$ 165 & 2198 $\pm$ 41 \\ 
\hline
40 & 4462 $\pm$ 59 & 10939 $\pm$ 157 & 3072 $\pm$ 53 \\
\hline
50 & 6943 $\pm$ 85 & 15844 $\pm$ 140 & 4062 $\pm$ 37  \\
\hline
60 & 9878 $\pm$ 150 & 21129 $\pm$ 311 & 4970 $\pm$ 79 \\
\hline
70 & 14316 $\pm$ 116 & 29420 $\pm$ 239 & 6449 $\pm$ 79 \\
\hline
80 & 17780 $\pm$ 201 & 35370 $\pm$ 306 & 7287 $\pm$ 59 \\
\hline
90 & 22736 $\pm$ 236 & 43799 $\pm$ 475 & 8484 $\pm$ 108 \\
\hline
100 & 27602 $\pm$ 221 & 51839 $\pm$ 378 & 9415 $\pm$ 70 \\
\hline
110 & 33176 $\pm$ 412 & 60820 $\pm$ 677 & 10443 $\pm$ 97 \\
\hline
120 & 39972 $\pm$ 501 & 71618 $\pm$ 868 & 11614 $\pm$ 142 \\
\hline
130 & 47841 $\pm$ 176 & 83651 $\pm$ 342 & 12544 $\pm$ 117 \\
\hline
140 & 54893 $\pm$ 661 & 95233 $\pm$ 1179 & 13829 $\pm$ 222 \\
\hline
150 & 64050 $\pm$ 526 & 109482 $\pm$ 919 & 15059 $\pm$ 136 \\
\hline
200 & 113727 $\pm$ 702 & 184621 $\pm$ 1179 & 19951 $\pm$ 162\\
\hline
250 & 190643 $\pm$ 1416 & 300888 $\pm$ 2327 & 25500 $\pm$ 200 \\ 
\hline
\end{tabular}
\caption{\label{tab:table1}The gate counts for the 
sparse SYK with $k=8.7$ in terms of two-qubit CNOT gates, Clifford group 
gates i.e., $\rm{H}, \rm{CNOT}$ and 
$T$-gates and just $T$-gates
for $N \le 250$ using first-order Trotter approach. 
The $T$-gate count can be further optimized but we do not discuss it here. 
The average gate count is computed over at least ten instances of the sparse
model ($k=8.7$). The two-qubit costs for the dense SYK model was 
discussed previously in Ref.~\cite{Asaduzzaman:2023wtd}.
} 
\end{table} 

\section{Summary} 

We have computed the gate complexity to simulate the sparse SYK model on near-term and fault-tolerant devices by calculating the
number of two-qubit gates and Clifford+$T$ gates required using the Trotter approach following the idea of decomposing the Hamiltonian into a cluster of terms where all terms commute in a given cluster. Once such clusters are constructed, each cluster is transformed to the computational basis
($Z$-basis). The resource estimates show that while dense SYK is not \emph{easy}, controlled sparsification with $k < 10$ will be possible
to study in the coming decade on quantum devices. We did not compute any observables
but simply focused on cost estimates in this work. In addition to the uniform sparsification
considered in previous works and here, there are possibly other effective ways of sparsifying the Hamiltonian based on the commutativity of the Pauli string representation of terms
rather than just random deletion such that complexity can be minimized without loss
of any holographic feature. However, reducing the Trotter complexity is not the complete story in
studying quantum chaos. Even with reduced complexity for the time evolution, the calculation of the
Lyapunov exponent for some $\beta\mathcal{J} \gg 1$ might still be difficult due to the challenges of
computing four-point correlations over prepared thermal quantum states.  

The extent to which the SYK model can be sparsified is promising because this hints at the fact that there might be other quantum many-body systems that admit holographic features and for which the Hamiltonian simulation also scales favorably in the degrees of freedom. It would be interesting to look for such models
and study them in addition to the sparse SYK model on quantum hardware in the coming decades. 
This is \emph{much easier} than other 0+1-dimensional holographic models, such as the BFSS (Banks-Fischler-Shenker-Susskind) matrix model ~\cite{Banks:1996vh} with large gauge groups that require resources (estimated by the number of logical qubits and the number of gates) much higher than required for factoring 2048-bit RSA integers \cite{Jha:2024}. 

\subsection*{Acknowledgements}
The research is supported by the U.S. Department of Energy, Office of Science, National Quantum Information Science Research Centers, Co-design Center for Quantum Advantage under contract number DE-SC0012704 and by the U.S. Department of Energy, Office of Science, Office of Nuclear Physics under contract DE-AC05-06OR23177. The author would like to thank ORF, ORD, and SJO for assistance during the completion of this work. The code to generate the $\textsc{Open QASM}$ circuits for the Hamiltonian simulation of SYK-type models using the Trotter method can be obtained from the author.

\bibliographystyle{utphys}
\raggedright
\bibliography{v2.bib}

\end{document}